\documentclass[conference]{IEEEtran}
\usepackage[a4paper,left=3cm,right=2cm,top=2.5cm,bottom=2.5cm]{geometry}
\usepackage{cite}
\usepackage[pdftex]{graphicx}
\usepackage[cmex10]{amsmath}
\usepackage{amssymb}

\usepackage{times}
\usepackage{array}
\usepackage[tight,footnotesize]{subfigure}
\usepackage[font=footnotesize]{subfig}
\usepackage{graphicx}
\usepackage{mathtools}
\usepackage{amsmath}

\def\intersect{{\;\cap\;}}

\usepackage{alltt}
\usepackage{shadethm}
\newshadetheorem{comment}{Comment}

\def\boxx{{\vcenter{\vbox{\hrule height.3pt
          \hbox{\vrule width.3pt height6pt
          \kern6pt\vrule width.3pt}\hrule height.3pt}}\;}}

\def\impos{{\;\vcenter{\hbox{\rule{5mm}{0.2mm}}} \vcenter{\hbox{\rule{1.5mm}{1.5mm}}} \;}}

\def\lrarrow{\leftrightarrow \kern-8pt \rightarrow}

\def\2{\frac{1}{2}}

\def\beq{\begin{eqnarray}}
\def\eeq{\end{eqnarray}}
\def\2{\frac{1}{2}}

\def\CAND{\;{\bf AND}\;}

\def\lrarrow{\leftrightarrow \kern-8pt \rightarrow}

\def\frightarrow{\rightarrow \kern-11pt /~~}
\def\reducesto{\simeq \kern -3pt >}
\def\intersection{\cap}

\usepackage{fancyhdr}					
\begin{document}
\newcommand{\strust}[1]{\stackrel{\tau:#1}{\longrightarrow}}
\newcommand{\trust}[1]{\stackrel{#1}{{\rm\bf ~Trusts~}}}
\newcommand{\promise}[1]{\xrightarrow{#1}}
\newcommand{\revpromise}[1]{\xleftarrow{#1} }
\newcommand{\assoc}[1]{{\xrightharpoondown{#1}} }
\newcommand{\rassoc}[1]{{\xleftharpoondown{#1}} }
\newcommand{\imposition}[1]{\stackrel{#1}{\impos}}
\newcommand{\scopepromise}[2]{\xrightarrow[#2]{#1}}
\newcommand{\handshake}[1]{\xleftrightarrow{#1} \kern-8pt \xrightarrow{} }
\newcommand{\cpromise}[1]{\stackrel{#1}{\frightarrow}}
\newcommand{\policy}{\stackrel{P}{\equiv}}
\newcommand{\field}[1]{\mathbf{#1}}
\newcommand{\bundle}[1]{\stackrel{#1}{\Longrightarrow}}

\title{Information and Causality in Promise Theory}
\date{\today}
\author{Mark Burgess\\~\\ChiTek-i AS\\~}
\maketitle
\IEEEpeerreviewmaketitle

\renewcommand{\arraystretch}{1.4}

\begin{abstract}
  The explicit link between Promise Theory and Information Theory,
  while perhaps obvious, is laid out explicitly here. It's shown how
  causally related observations of promised behaviours relate to the
  probabilistic formulation of causal information in Shannon's theory,
  and thus clarify the meaning of autonomy or causal independence, and
  further the connection between information and causal sets. Promise
  Theory helps to make clear a number of assumptions which are
  commonly taken for granted in causal descriptions. The concept of a
  promise is hard to escape. It serves as proxy for intent, whether a
  priori or by inference, and it is intrinsic to the interpretations
  of observations in the latter.
\end{abstract}



\section{Introduction} 

Promise theory describes interactions between generalized agents
and their possible outcomes. Promises declare possible causal
pathways by defining and documenting `outcomes' as possible boundary
states of process graphs. In so doing, they provide scalable
definitions of `intent without anthropomorphism' and measurement, based on agents' assessments
of one another. Promise Theory has now been in use for over 15 years
and has been applied to many different kinds of process networks\cite{promisebook,treatise2}.

The goal of this letter is to describe the relationship between
Promise Theory's model of agents and promises, and the statistical
information passed between them as described by Shannon's Theory of
Communication.  Information Theory relies on scale invariant
probabilities, whose meanings are inherently ambiguous, but sometimes
phenomena are scale dependent. One of the aims of Promise Theory is to
move beyond these ambiguities, while suppressing details that are not
measurable in practice anyway for many systems.

\section{Problem and notation}

Using the standard promise notation from \cite{promisebook}, 
there are three main interaction patterns we need to distinguish and account for:
Let $A$ be any agent (on any scale), whose interior structure is unspecified.
In this letter, its identity will be closely associated with its role in
a particular interaction pattern so we can simplify the notation by taking
$A \in \{S,R,I\}$, for sender, receiver, and intermediate nodes.

Consider three basic cases. The simplest is a unilateral promise declaration
of $b_S$ by $S$, without acceptance by a promisee $R$:
\beq
S ~~~~ \promise{+b_S} R,
\eeq
i.e. the promise meets with `deaf ears'.
The second is a promise of $b_S$ with partial or complete acceptance $b_R$ by its recipient:
\beq
S ~~~
\begin{array}{c}
 \promise{+b_S} \\
 \revpromise{-b_R} 
\end{array}
R,
\eeq
and the final is the chain propagation of influence by conditional promises,
where $b_I$ is promised if and only if $b_S$ is accepted by $I$:
\beq
S~~~~
\begin{array}{c}
 \promise{+b_S} \\
 \revpromise{-b_I} 
\end{array}
I
\begin{array}{c}
 \promise{+b_I | b_S} \\
 \revpromise{-b_R} 
\end{array}
R 
\eeq 
Since agents make promises only for themselves, these labels $S$, $I$,
and $R$ play a second role as subscripts to indicate the source of each promised
measure.  The promise bodies represent the details of what is
promised, and these are set-valued measures.

\begin{figure}[ht]
\begin{center}
\includegraphics[width=6cm]{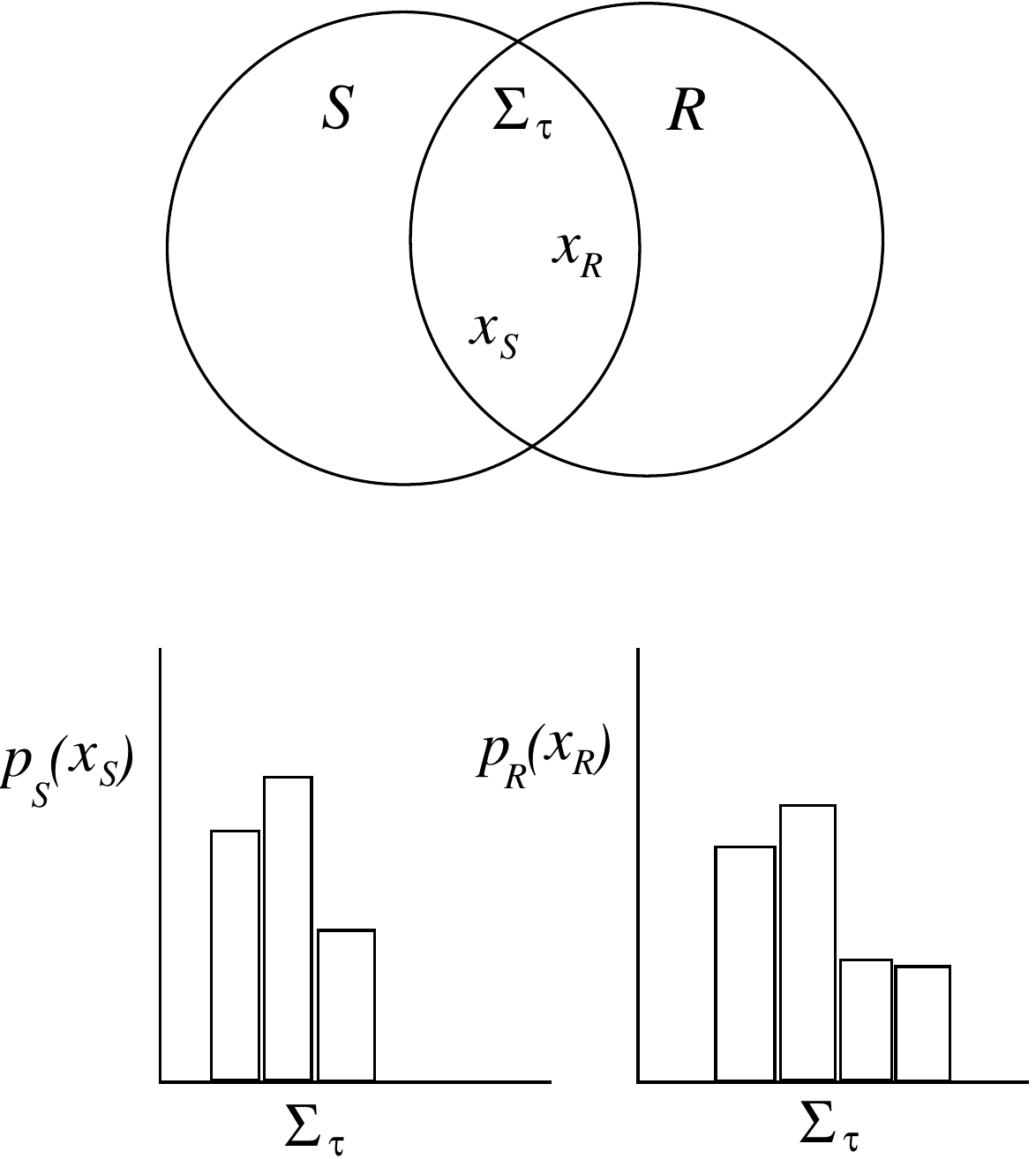}
\caption{\small In Shannon's theory, information is defined
  statistically, from transmissions composed from prior alphabets
  $\Sigma_A$ for sender ($A=S$) and receiver $A=R$. The entropic
  measures of interaction `effectiveness' follow from statistical
  observation and inference between the simultaneous observations of
  these endpoint agents $S$ and $R$. The common view of this gets into trouble once
  more careful relativistic considerations are taken into
  account.\label{inf1}}
\end{center}
\end{figure}

An agent $A'$ must assess the extent to which a promise $\pi_A$ made by $A$ has been
kept or not.  If privy to the interior process states $\Sigma_A$ for a
promised process, this is partly determined by that information, else
it's arbitrary.  The assessment by $A'$ is denoted $\alpha_{A'}(\pi_A)$, and
can take several forms, some of which may involve `probabilities' for certain
states $\sigma_i \in \Sigma_A$, where the Latin indices run over the
different symbols in $\Sigma_A$ (see figure \ref{inf1}).
The symbols, which are members of the body sets $b_A$ take values from the
alphabet of interior agent states $\Sigma_A$, and I assume that every agent speaks a language
composed of an alphabet $\Sigma_\tau$ for that particular promise type $\tau$\footnote{A further subtlety, which one normally ignores based on an assumption of
homogeneity or global symmetry for types $\tau$, is that the promise body belongs uniquely
to its originating agent $A$, so the alphabets for a promise of type $\tau$
are really private to each agent--and we should really write $\tau_A$.
However, since non-shared types and alphabets would not result in binding at all,
we can effectively ignore those cases where agents speak incompatible languages
and absorb such cases into the assessments of promises that are not kept.}.
  For a promise: 
\beq 
\pi: A \promise{b_A} A', 
\eeq 
the promise body $b_A$ is a constraint on $A$, consisting of
a type label $\tau$ and a measure which belongs to the alphabet of the
promise language $\Sigma_\pi$ 
\beq 
b = \langle \tau, \chi \in \Sigma_\pi \rangle 
\eeq 
The overlap of two promise bodies of the same
type of promise\footnote{For example, and agent $S$ could be an Light Emitting Diode (LED) that
promises from within a finite alphabet of red, green, blue (RGB)
symbols. The receiver might be a light sensitive detector which can
only detect shades of what it calls green (G).}. Finally, we recall that agents
of any type exist in both positive and negative polarities.

An assessment of whether or not a promise is considered kept 
may be made on a variety of scales and criteria. 
The symbols $\alpha_A(\cdot)$ is used for assessments of various kinds, to be detailed
in context. The semantic or symbolic assessment of a single sample, for each single promise-keeping event 
is simple the outcome:
\beq
\alpha_A(x(\pi)) :   \pi \rightarrow (x\in\chi_\tau)
\eeq
where $x(\pi)$ denotes the sampling of a symbol $x$ from the channel
formed maintained by the keeping of the promise $\pi$, Alternatively,
we could evaluate the semantic average assessment, relative to the
promise declaration:

\beq
\alpha_A(\pi) :   \pi \rightarrow \{ \text{\sc kept}, \text{\sc not-kept} \}.
\eeq

A collection of such events leads to a distribution of outcomes, which we can
denote either as an average $\overline\alpha()$ or as a pro-forma `probability'\footnote{The concept
of a probability involves plenty of semantics that are often taken for granted. Here we needn't
take issue with different definitions, as any will do the job.}:
\beq
\left. \begin{array}{c}
\overline\alpha_A(\pi) \\
p_A(\sigma_\pi)   
\end{array}
\right\}
: \sigma_\pi\in\chi_\tau \rightarrow [0,1]
\eeq
where the ensemble is defined over a specified set of $S$ samples
\beq
\sum_{i=1}^{N} \frac{\alpha_A(x(\pi) = x)}{N} \rightarrow p_A(x)
\eeq

These `probabilities' are the quantitative scale ratios, used in
definitions of information, according to the Shannon theory of
communications\cite{shannon1,cover1}.  How ensembles are constructed
is important, but not defined a priori. If one has a controlled environment
which can promise repeatable configurations, them there can be spacelike (co-temporal) or
frequentist probabilities, and there are timelike (temporal) of
Bayesian probabilities, which have different interpretations. In
either regime, we have probabilities $p_A$ assessed by each agent $A$.
However, in order to get to information, we need assessments made by
more than one agent: both a sender and a receiver.

We can use these measures to pursue three issues:
\begin{itemize}
\item The meaning of autonomy, or causal independence of agents.
\item The transmission of intent or expectation as symbols.
\item The transmission of observations and assessment from symbols.
\end{itemize}
These three matters are related but distinct.  In related work,
considering the concept individuality \cite{sfiinfo}, the authors use
mutual information as the criterion by which to define autonomy or
causal independence of agents. They show that the assumption of
individuality is consistent with immunity from external information
propagation. A compatible answer is implicit in Promise Theory, but
without the implicit assumptions about probability.  Here, the axioms
contend that all agents are a priori autonomous or causally
independent, and may forego that autonomously, which amounts to a subtle difference. The two viewpoints are
entirely consistent where they overlap, but the formulation
based on Shannon's mutual information is not
relativistically covariant, whereas a formulation based on promises
is. In principle, the promise view is not only simpler but reveals more of the interaction
picture than the information view, since the entropy functions are
based on ensemble averages. I'll return to this at the end.

\section{Simultaneous characteristics}

\begin{figure}[ht]
\begin{center}
\includegraphics[width=5.5cm]{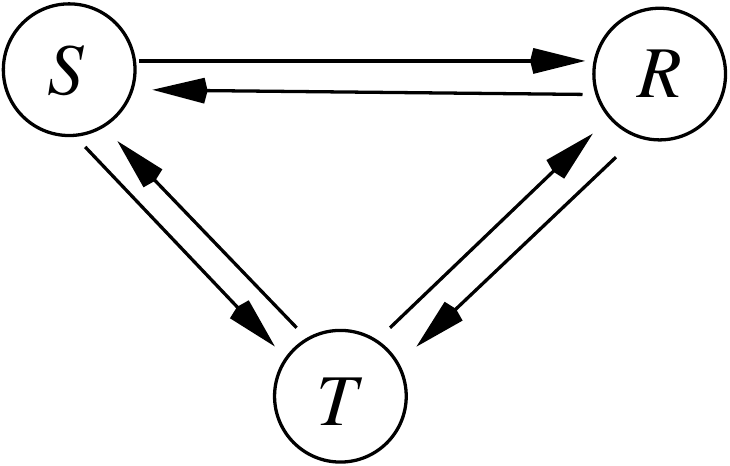}
\caption{\small Observation by a trusted third party observer.\label{inf2}}
\end{center}
\end{figure}

In Shannon's statistical measures of information, there is no
relativity of end points incorporated into the picture, despite the
end points being causally distant from one another. Events are assumed
simultaneous and therefore form local matrices as viewed by a single
`godlike' observer, with infinite and immediate access (figure
\ref{inf2}). Promises remain true to a
tensorial picture with explicit construction of multi-local behaviours.

A joint probability matrix incorporates the idea that the receiver may get
a transmitted symbol wrong with a certain off-diagonal probability, if the
non-diagonal elements are non-zero. 
The joint probability matrix, which depends on two ends of a causal channel, is
defined by:
\beq
p_A(\sigma_S,\sigma_R) \equiv p_A(\sigma_S \CAND \sigma_R), 
\eeq
as used in Shannon's formulation of informational entropy,  is 
assumed to be observable (by the godlike observer who computes information
transfer). It reduces to the lower rank product state $p_A(\sigma_S)p_A(\sigma_R)$ when $\sigma_S$ and $\sigma_R$ are
independent variables. As we see below, the existence of this measure assumes
coarse graining in a time and space, so its interpretation may be ambiguous.
Independently of the definition of measures, the joint matrix 
represents the agreed `probability' that $S$ and $R$ sample common symbols at the same
moments. Since the symbols are actually private 
interior representations of the agents, in Promise Theory, 
they are assessed autonomously.
What this really means is that the observation of a correspondence of symbols at the two ends of the channel 
$\sigma_S \sim \sigma_R$, when the symbols coincide in
the overlap $\chi = \chi_S \intersection \chi_R$, are therefore `equal' in the sense that
they play an invariant role which is understood (independently) by each end of the
channel between $S$ and $R$. The agents needn't have a common representation
of these symbols, so we cannot say that $\sigma_S = \sigma_R$, as measured by a godlike third party,
but they consistently represent the
same information to each party.

\section{Scalar promise (case 1)}

An assessment of probability for interaction by any agent (of
symbols $\sigma \in \Sigma_\pi$ for the promise $\pi$), over some ensemble
of multiple promise-keeping events, relies on repeated observations under coarse-grained
conditions.
In the case of a unilateral promise from $S$, {\em with no acceptance} by
$R$, $R$ makes no assessment of any promise $\pi$ at all (indeed, it
doesn't even know about it), so we have simply: 
\beq
p_S(\sigma_S,\sigma_R) &=& p_(\sigma_S)\times 0\\
p_R(\sigma_S,\sigma_R) &=& 0 \times p(\sigma_R) 
\eeq 
i.e. the joint probability is
entirely separable, and is identically zero, as the agents are fully independent. Any similarity
of symbols, seen only by a privileged third party $T$ (figure
\ref{inf2}), they were able to measure would be entirely
coincidental, as they are not able to observe one another. So the
mutual information of the agents is identically zero in this case:
\beq I(S;R) = 0.  \eeq

Before leaving this elementary case, we should point out the concept
of scope, which is a shorthand for a large number of implicit
promises.  
\beq 
\pi: S \scopepromise{+b_S}{\Omega} R.  
\eeq 
The symbol $\Omega$ for scope, implies a set of agents to whom the
promise of observability has been granted and accepted. In this case, only $R$
has been offered this information, but this is important for the next two cases. In other
words, only agents $A \in \Omega$ can form assessments of the promise
$\pi$, even though they are not explicitly mentioned in it, and the
outcome may not be any of their business per se.  These agents in
scope are the typical observers in quantum mechanical scenarios, for
instance. They are third parties who look upon other agents and
calibrate the outcomes according to their own alphabets of states.

\section{Promise binding (case 2)}

From the foregoing definitions, we can now complete the promises for an information channel,
in the Shannon sense, for a single promise with $b(\pi) = \langle \tau, \chi \rangle$, by adding 
the acceptance promise, which is normally taken for granted: 
\beq
S &\promise{+b_S}& R\nonumber\\
R &\promise{-b_{RS}}& S,\label{binding} 
\eeq 
Here, it's assumed that
$\tau(b_S)=\tau(b_R)$, i.e. the agents are aligned in their `intent'.
Now that $R$ accepts the promise from $S$, \beq
b_S &=& \langle \tau, \chi_S \rangle\\
b_{RS} &=& \langle \tau, \chi_R \rangle \eeq and, for a non-zero
channel to form, we require the channel overlap of languages to be
non-empty: \beq \chi_{R|S} = \chi_S \intersection \chi_R \not=
\emptyset.  \eeq

The subscripts here seem pedantic, but are used to remind us that what
is offered $b_S$ is initiated only by $S$ and only concerns $S$'s
interior state (call it $\sigma_S$); and, in receiving data from $S$,
$R$ has its own a choice about what it is willing or able to
accept---that choice needs an `$R$' annotation to identify it as $R$'s
`intent'. Even though it's based on data from $S$, the data are only
accepted at $R$'s behest. It is not necessary to assume that $R$  is compelled
to accept, by some force. Indeed, we know that this viewpoint is inconsistent\cite{promisebook}.

Using the subscript $R|S$ ($R$'s acceptance given $S$'s offer), we
indicate this implicit causal dependency. Later (in case 3) this
dependency will be made explicit when an agent bases a new promised
value (sent downstream) on a prior one received from upstream, forming
causal chains.

With channel binding now accepted, there is a new issue: for agents
to be able to assess one another they must be promised
access to information about each other's interior states, not only a
declaration of the promise of expected behaviour. In other words, 
the promises in (\ref{binding}) are sufficient to enable causal
influence, but are not sufficient to be able to
assess it from mutual information, since the promises in
(\ref{binding}) are not mutual, merely complementary.
This is related to the generally assumed observability issue (even for godlike observers).
We might treat observability promises as part of `scope' to avoid a proliferation of
promise arrows, but here that would be sweeping key issues under the rug. 
We have to deal with two kinds of promise in order to verify behaviours observationally:
\begin{itemize}
\item A promise of behaviour is something with the status of a `charge',
it allows inference of propagated influence.
It refers to  promises which are locally invariant over the events that
confirm them. 

\item A promise of observational outcome, on the other hand, requires
  access to private interior states, which is additional information
  about dynamical changes. and the observation of acceptance is an
  acknowledgment.

Let's define the alphabet of possible values generated by
the set of relevant interior states for $S$ and $R$ by
 $\sigma_S$ and $\sigma_R$. These states exist at opposite ends of the channel
but they can be subject of a promise made by each end to the other:
\beq
S &\promise{+\sigma_S}& R\\
R &\promise{+\sigma_R}& S
\eeq
In order to be observed, the other end must accept the promised
symbols, but they may accept a different set:
\beq
S &\promise{-\sigma_{S|R}}& R\\
R &\promise{-\sigma_{R|S}}& S
\eeq
so that what is actually possible to transmit is the overlap:
\beq
S \rightarrow R: \sigma_S \intersect \sigma_{R|S} \le \sigma_S\\
R \rightarrow S:  \sigma_R \intersect \sigma_{S|R} \le \sigma_R
\eeq
This is what we mean by observability. The $\le$ can also account for noise, but
it has a different semantic origin. Noise could always be corrected, as Shannon
showed, but the inability or unwillingness to receive certain symbols cannot.

These observability promises are often presumed as `bundled' in our world view, when the first kind 
of promise has been given, but that's not strictly necessary. We don't always have access
to observe the outcomes of events. For example, a particle might promise a certain charge,
but we have no exterior field or detector to register the forces it may experience.
\end{itemize}

Over one or more assessments by the two agents, ensembles
can be formed to define probability measures from each agent's
independent observational perspectives. 
\beq
p_S(\sigma_S,\sigma_R) \not= 0\\
p_R(\sigma_S,\sigma_R) \not= 0.
\eeq
$S$ assesses this as the result of the confirmation it
sent and the acceptance of an acknowledgment received.
The existence of this
matrix now depends on two sets of promise bindings: 
\begin{itemize}
\item Invariant intent (e.g. `charge' in physics), and
\item On-going observability of interior states (events and transitions).
\end{itemize}

When the alignment of states is merely coincidental, then over
ensemble we would observe that
\beq
p_S(\sigma_S,\sigma_R) &\rightarrow& p_S(\sigma_S)p_S(\sigma_R)\\
p_R(\sigma_S,\sigma_R) &\rightarrow& p_R(\sigma_S)p_R(\sigma_R),
\eeq
which contains non-local information.
This assessment cannot be made unless observability promises
have been offered, accepted, and kept for the agent assessing the joint
states.
The mutual information, as used in \cite{sfiinfo}, requires this
minimum observability for some observer. Assuming an agent $A$ has
such access, then it is the average overlap remaining once the
probability of coincidence of random processes is subtracted, as
measured by an independent observer $A$: 
\beq 
I_A(S;R) =
\sum_{\sigma_S,\sigma_R} \; p_A(\sigma_S,\sigma_R) \log
\frac{p_A(\sigma_S,\sigma_R)}{p_A(\sigma_S)p_A(\sigma_R)},\nonumber
\eeq which is clearly zero when\footnote{The weakness of treating the agents
  as random processes is that we have to deal with average causality
  and probable measures over coarse grained ensembles, when what we
  really want to deal with individual interactions from the bottom
  up.}:
\beq
p(\sigma_S,\sigma_R)=p(\sigma_S)p(\sigma_R).
\eeq

We can now consider which promises are required to evaluate the probabilities
presumed to be calculable in the informational entropies of a channel.
\begin{itemize}
\item A self-assessment $p_A(\sigma_A)$ (e.g. $p_S(\sigma_S)$) can always be obtained
`immediately' according to $S$'s interior process and clock. 
\item A remote assessment $p_A(\sigma_{A'})$ (e.g. $p_S(\sigma_R)$) is reliant on data
being propagated along the information channel, which implies that source and
receiver are never simultaneous. This transmission depends on two promises
being kept. For the example:
\beq
R &\promise{+\sigma_R}& S\\
S &\promise{-\sigma_{S|R}}& R.
\eeq
Looking for a measure of correlation between the symbols at source and
receiver still doesn't discount the possibility that random processes
led to a coincidence of symbols. For the causal information transmission
we want the conditional transmission only, which can be excluded statistically
over ensemble averages by using the definition for mutual information, but 
here I want to emphasize that this mutual information is only
an average proxy for causal propagation implicit (namely that $S$ signals
$R$ and then $R$ acknowledges receipt, as in so-called reliable network protocols, such as
TCP/IP):
\beq
S &\promise{+\sigma_S}& R\\
R &\promise{-\sigma_{R|S}}& S\\
R &\promise{+\sigma_R | \sigma_{R|S}}& S\label{det}\\
S &\promise{-\sigma_{S|R}}& R.\label{cond}
\eeq
This is the precise statement, on a transactional basis, which is usually coarse-grained 
to yield common
sets of mutual probabilities.
\end{itemize}
By stating (promising) the mutual information above, we effectively
timestamp all events on the interior of a single agent, e.g. $R$
claiming that this is instantaneous for $R$. This is an approximation which
is good enough on human scales, but which fails on computational and quantum scales.

On the question of autonomy, or causal independence: regardless of
our ability to define probabilistic measures, one sees that the 
promise in equation (\ref{det}) is empty if the conditional acceptance
of the input in (\ref{cond}) is absent, which is the only one in which
an agent is influenced by another agent. So, at a basic level, the agents
are always causally independent, but may promise to forego that autonomy
by accepting inputs from other agents. It's the presence of such receptor promises
which therefore represent the causal `boundary' for influence. This applies
on any scale, since we have not made any assumptions about the interior
nature of the agent. It's a form
of Gauss' law, noted in \cite{spacetime2}, which may be expressed by saying
that what is promised from within a boundary depends a priori only on what
is on its interior.

We also see these points reflected in the flow of process time.
In a classical Newtonian view, time is a universal and simultaneous quantity
that presumes instantaneous access to a single calibrated clock for the entire
universe. We know this to be an idealization that fails under many circumstances,
and we must instead specify which observer's clock is being used to count
events that we call time\cite{observability,smartspacetime}.
The only things that $S$ and $R$ know about observations of one another is
that their receipt comes after the samples were obtained. So, according to either
$S$'s clock $t_S$ or $R$'s clock,
\beq
t_S(\sigma_S) < t_S(\sigma_{R|S})\\
t_R(\sigma_S) < t_R(\sigma_{R|S}),
\eeq
and similarly,
\beq
t_S(\sigma_R) < t_S(\sigma_{S|R})\\
t_R(\sigma_R) < t_R(\sigma_{S|R}).
\eeq
This might seem excessively pedantic for mundane human systems,
or biological timescales, but these distinctions are quite important
for processes that race one another with split-second timings in computer networks,
and sub-atomic physics.

A third party observer watching such a transition from an independent vantage point
could calibrate as an impartial arbiter (see figure \ref{inf2}); however, its
ability to do so is only uncontested if the promises of observability of $S$ and $R$
by $T$ are much faster than the changes taking place between $S$ and $R$\footnote{This notion is
encapsulated in the Shannon-Nyquist sampling law. In classical science, we are used
to observing rather slow transitions using light signals which are very fast, so these
matters become negligible.}.
If we try to parameterize the separation between these causal
interactions, then we have a choice about how to represent the partial
ordering. Space and proper time can be proxies for that ordering, but
since the separation is only a convolution of two autonomous
processes, the interpretation is moot.  The joint probability has the
form of a faithful assessment by a sufficiently fast third party
(where `fast' means satisfying the Nyquist law over relevant
timescales for $\pi$):

\beq
&~& p(S,R) =\nonumber\\
&=& \alpha_T\left( \pi^{(+)}(S) \CAND \pi^{(-)}(R)\right) \nonumber\\
&=& \alpha_T\left( \pi^{(+)}(S)\right) \cdot \alpha_T\left( \pi^{(-)}(R)\right)\nonumber\\
&=& \int_0^\infty ds \left( \alpha_T^{(+)}(S,t_T) \cdot \alpha_T^{(-)}(R,t_T+ds)\right)\nonumber\\
\eeq
If we assume a vanishing observability delay, then this has the simplified form:
\beq
p(+,-) &=& \int_0^\infty dA \left( \psi^{(+)}(A) \psi^{(-)}(A)\right)
\eeq
which has the form of a spatial convolution, familiar in quantum mechanics\footnote{In a Hilbert
space model, which can assume lossless probabilities of a closed system, the + and - are natural
Hermitian conjugates.}.

\section{Causal chains (Case 3)}

All the elements are now in place for the general case of a causal
chain. Note, there is no assumption of a Markov chain, without hysteresis, nor any
presumption of a global symmetry. Agents may contain any amount of
memory and can incorporate hysteresis, absorption, and so on. They also
retain their individuality by default, and must make explicit promises
to constrain their behaviour and demonstrate transmission of influence.
The question is whether such effects play
a prominent or a negligible role in the observed outcomes.

The main difference in between equation (2) and equation (3) is that
there is now a promise with an explicit dependence of a prior outcome
from another promise, which is propagated from one agent to another.
This was implicitly present in the acknowledgment of observable
transmission, but it also carries over into successive interactions
(e.g. `collisions' or transmission relays).
All the elements are therefore in place, from the discussion
of observability, to cover these interactions.

There are two things to note: first, receptor promises ($-b_A$) are
the key to transmission of influence. Influence is not conferred
automatically by emission of a signal alone---absorption is an
autonomous behaviour too.
Next, the foregoing implies that intermediate agents play a key role in forming barriers
to transmission (e.g. we can think of the purpose of vaccines to stem disease transmission),
as long as they don't make an independent promise to accept and relay information from a prior
agent in the chain with high fidelity. Intermediate agents therefore also serve as the elementary
construction for modelling noise and other environmental input channels within a
fundamental system. The intermediaries may be:
\begin{itemize} 
\item Intentionally bad actors.
\item Faulty, low fidelity replicators.
\item In possession of covert channels that accept information from other sources in the `environment' $E$:
\beq
I \promise{-b_E} E.
\eeq
Such influences would impinge on the relay function for propagation from $S$ to $R$ if the promise in (3)
were replaced by
\beq
I \promise{+b_I \;|\; b_s,b_E}.
\eeq
\end{itemize}

The relationship between $b_S$ and $b_I$ is of crucial interest in determining faithful
propagation along a chain. From a state of initial causal independence, there is no
reason why $S$ and $I$ should make the same promise, unless they have been calibrated
by a `global symmetry' pertaining to all agents. Although such global symmetries
exist in nature, from particle physics to biology, the reason for them is unclear
and they seems to violate local relativity principles. All such agents appear to have a common origin,
or promise to accept causal influence from a common source (see the matroid pattern in \cite{promisebook}):
\beq
S &\promise{-b_M}& M\\
I &\promise{-b_M}& M
\eeq
and
\beq
S &\promise{+b_S | b_M}& A\\
I &\promise{+b_I | b_M}& A'.
\eeq
In biology, this is indeed the case, where cells are formed by replication from a 
single source. In particle physics, we simply don't know what underlying information
is at work.

\section{Summary}

Comparing Shannon's statistical theory of communication, commonly
known as Information Theory, we find that Promise Theory's simple partial ordering both
simplifies the notations, suppressing probabilities, and thus eliminates
the need to define statistical ensembles, with all the attendant
assumptions therein.  The probabilistic formulae for the various
informational entropies are based on statistical ensembles and
observations, but the causal concepts are available on a deeper level,
as we can see explicitly in Promise Theory. The promise formulation
shares something with formalisms like Quantum Field Theory; this is to be
expected, given the prominence of agent relativity. The Shannon theory is
an absolute spacetime theory, which is adequate for its technical origins,
but which is unsuitable for more general processes.

In order for agents to give up their autonomy and become a completely
deterministic game piece, within a ballistic model of motion, like a
Newtonian system or a Markov chain, they have to be calibrated to the
same identical promises (spacetime homogeneity), and they have to completely
forego individual autonomy (locality).
{\em A priori} autonomy, or causal independence, is an axiom in Promise Theory, represented
by the need for (-) polar promises. This
is the natural `bare' state of any agent. The addition of promises may
then constrain agents to cooperate. This is consistent with the
conclusions reached in \cite{sfiinfo}, which introduce probabilistic
considerations, perhaps unnecessarily. The reason is clearly due to the underlying causal ordering
implicit in the information channel. This model may also be seen as a deeper
structural explication of the Causal Set spacetime model developed for discrete spacetime
structure \cite{myrheim1,sorkin1,surya1,dowker1,spacetime4}.

Causal independence is not really a pervasive property of an agent,
but rather it's a property of each promise made by an agent. Certain
promises may be entirely determined by exterior influence (or at least
be indistinguishable from being entirely determined to an external
observer), whereas others may appear entirely autonomous. This has as
much to do with promise observability as autonomy---which, indeed, is the
curse of relativity: being trapped within the rules of the system one aims to observe.

\bibliographystyle{unsrt}
\bibliography{spacetime}

\end{document}